\documentclass[8pt, review]{elsarticle}

\immediate\write18{%
makeindex -s nomencl.ist -o \jobname.nls -t 
\jobname.nlg \jobname.nlo%
}

\usepackage{subfig}
\usepackage[inkscapelatex=false]{svg}
\usepackage{afterpage}
\usepackage{pdflscape}
\usepackage{longtable}
\usepackage{comment}
\usepackage{fullpage}
\usepackage[hyperfootnotes=false, hidelinks]{hyperref}
\journal{xxx}
\usepackage{eurosym}
\usepackage{xcolor}
\usepackage[per-mode=symbol]{siunitx}\usepackage[version=4]{mhchem}
\usepackage{bm}
\DeclareSIUnit{\EUR}{\text{\euro}}
\usepackage{url}
\usepackage{floatrow}
\usepackage[toc,page,titletoc]{appendix}
\usepackage[capitalise,nameinlink]{cleveref}
\usepackage{amsmath}

\usepackage{booktabs} 
\usepackage{tikz}

\crefname{supp}{Supplement}{Supplements}

\usepackage{framed} 
\usepackage{multicol} 

\usepackage[noprefix]{nomencl} 
\makenomenclature
\renewcommand*\nompreamble{\tiny \begin{multicols}{2}}
\renewcommand*\nompostamble{\end{multicols}}

\def\sectionautorefname{Section}
\makeatletter
\def\ps@pprintTitle{\ps@empty}
\makeatother

\biboptions{numbers,sort&compress}

\begin{document}

\def\sectionautorefname{Section}
\def\subsectionautorefname{Subsection}
\def\subsubsectionautorefname{Subsection}

\begin{frontmatter}
\title{Revealing design archetypes and flexibility in e-molecule import pathways using Modeling to Generate Alternatives and interpretable machine learning}

\author[UCL]{Mahdi Kchaou\corref{mycorrespondingauthor}}
\ead{mahdi.kchaou@uclouvain.be}

\author[UCL]{Francesco Contino}
\author[UCL]{Diederik Coppitters}

\cortext[mycorrespondingauthor]{Corresponding author}

\address[UCL]{Institute of Mechanics, Materials and Civil Engineering (iMMC), Universit\'e catholique de Louvain (UCLouvain), Place du Levant, 2, 1348 Louvain-la-Neuve, Belgium}

\begin{abstract}
Given the central role of green e-molecule imports in the European energy transition, many studies optimize import pathways and identify a single cost-optimal solution. However, cost optimality is fragile, as real-world implementation depends on regulatory, spatial, and stakeholder constraints that are difficult to represent in optimization models and can render cost-optimal designs infeasible. To address this limitation, we generate a diverse set of near-cost-optimal alternatives within an acceptable cost margin using Modeling to Generate Alternatives, accounting for unmodeled uncertainties. Interpretable machine learning is then applied to extract insights from the resulting solution space. The approach is applied to hydrogen import pathways considering hydrogen, ammonia, methane, and methanol as carriers. Results reveal a broad near-optimal space with great flexibility: solar, wind, and storage are not strictly required to remain within 10\% of the cost optimum. Wind constraints favor solar-storage methanol pathways, while limited storage favors wind-based ammonia or methane pathways.

\end{abstract}


\end{frontmatter}



\section{Introduction} 
\label{Introduction}

The EU aims to import 10 million tonnes of hydrogen by 2030 to meet its carbon-reduction target~\cite{EU_2030_target}. The general vision is that hydrogen will be produced in remote, renewable-rich regions using electrolysis powered by wind or solar~\cite{vilbergsson2023can}. Because transporting pure hydrogen is challenging, it can be converted into carriers such as ammonia for shipping or pipeline delivery~\cite{verleysen2023build,roos2021cost}. Once imported by the densely populated, industrialized area, hydrogen has the potential to serve multiple roles, from long-distance energy transport and seasonal storage to the decarbonization of heavy industry such as steel, chemicals, and cement~\cite{ishaq2022review}.

Several studies have examined the techno-economic aspects of producing and transporting hydrogen-based energy carriers~\cite{brandle2021estimating}. The Hydrogen Import Coalition~\cite{coalition2021shipping}, for example, assessed imports of renewable energy via hydrogen carriers shipped from various locations to Belgium, demonstrating both technical feasibility and cost-effectiveness. At a small scale (transport distances of 250 km), Dias~et~al.~\cite{V.Dias} showed that among four carriers: hydrogen, ammonia, methane, and methanol, hydrogen is the least favorable option for a production, transport and reconversion to electricity chain. In contrast, the other carriers are more cost-competitive, with methanol emerging as the cheapest option. This is largely due to their easier handling and the availability of existing infrastructure, as is the case for methane. At a larger scale, Hampp~et~al.~\cite{hampp2023import} evaluated imports of e-chemicals from diverse regions to Germany. They found that all e-chemicals could be imported at lower cost than producing them domestically. However, when meeting hydrogen demand directly, importing hydrogen was more attractive than importing carriers and converting them upon arrival. Other comparative studies suggest that ammonia is the most economical at large scales, while methanol can be competitive at small scales if low-cost CO$_2$ is available~\cite{CUI202315737}. Johnston~et~al.~\cite{JOHNSTON202220362} further showed that transporting ammonia by ship is the cheapest option for both Tokyo and Rotterdam. Yet results are not uniform: analyses of imports from North Africa and South America indicate that, under some assumptions, importing hydrogen may be more costly than domestic production~\cite{GALIMOVAfirst,GALIMOVAsecond}. Overall, these findings underscore that conclusions depend strongly on techno-economic assumptions and case-specific conditions.

A common limitation of these studies is that they rely on deterministic parameters~\cite{coppitters2023energy4}. In reality, both technical and economic inputs are subject to uncertainties: e.g. solar irradiance ($\pm 3.11\%$ in Morocco~\cite{verleysen2023build}), technology efficiencies (from $0.532$ to $0.648~\text{kWh/kg}_{\text{NH}_3}$ for the Haber–Bosch unit~\cite{verleysen2023build}), and investment costs (from $455$ to $5317 \text{\EUR/}\text{kW}$ for the Haber–Bosch unit~\cite{verleysen2023build}). Such uncertainties can affect system performance~\cite{mavromatidis2018review}. To account for this, many studies conduct sensitivity tests, varying one assumption at a time~\cite{hampp2023import}. While useful, this approach cannot fully capture how uncertainties propagate through models and simultaneously impact the performance of the system, including synergistic and antagonistic interactions. For this purpose, systematic Uncertainty Quantification (UQ) is required. To assess the importance of parametric uncertainty, Coppieters~et~al.~\cite{COPPITTERS2019310} used the polynomial chaos expansion method to quantify uncertainties in a coupled PV-electrolyzer system. They reported a robust Levelized Cost of Hydrogen (LCOH) of 6.4 \EUR/kg in Johannesburg, with a standard deviation of 0.74 \EUR/kg, while the discount rate was identified as the main driver of cost variation. In a similar hydrogen production context, Wolf et al.~\cite{wolf2024levelized} applied uncertainty quantification to estimate the 2050 LCOH in Northern Africa and Europe, and found large uncertainty ranges: PV-based hydrogen in Spain varied from about 1.66 \EUR/kg to 3.12 \EUR/kg, and in Morocco and Algeria from about 1.70–1.80 \EUR/kg up to 3.2–3.3 \EUR/kg. 

In addition to parametric uncertainties, structural uncertainties linked to decision-making play a critical role~\cite{korem2025modeling}. Unlike parameter uncertainty, which assumes the model itself is correct, real projects also face uncertainty about goals, system boundaries, and stakeholder preferences. These hard to quantify uncertainties can, in practice, fundamentally alter what is considered optimal in theory. Examples from past renewable projects illustrate these effects. Social resistance has delayed or canceled developments, such as the Nevada PV farm opposed by local communities~\cite{pv_nevada_fail}. Technological limits for large scale batteries, like rate capability, lifespan, and raw material needs, continue to hinder their deployment despite favorable techno-economic projections~\cite{cambridge_battery_lim}. Policy shifts, such as Morocco’s move from Concentrated Solar Power to PV coupled with storage~\cite{Noor2024}, can redirect investment away from designs once seen as optimal. Legal disputes, as with wind projects in Spain~\cite{wind_spain_fail}, further illustrate how institutional processes can reshape or halt projects. 

For hydrogen import pathways, these uncertainties are especially important. Choices about carriers, the balance between imports and domestic production, or the expansion of ports and pipelines depend not only on costs but also on politics, lobbying, regulation, and public acceptance. Geopolitical analyses show that international relations and energy security concerns can determine which routes are viable, even when cheaper options exist~\cite{pflugmann2020geopolitics}. In Europe, lobbying by industry groups has been shown to influence hydrogen policy and favor certain pathways~\cite{flath2025lobbying}. Public acceptance also matters: survey evidence~\cite{palanca2025philippines} shows that awareness, trust, and local concerns can delay or block hydrogen projects. Thus, what appears optimal in techno-economic models may prove politically unworkable, socially resisted, or institutionally blocked.

A way to address non-quantifiable or structural uncertainties in energy system modeling is Modeling to Generate Alternatives (MGA). Rather than searching only for the lowest-cost solution, MGA identifies near-optimal configurations that differ in design, thereby allowing qualitative, institutional, or acceptance criteria to be considered when selecting among them~\cite{Brill1979}. DeCarolis~\cite{decarolis2011} illustrated the application of this methodology to energy systems, showing how maximally different yet cost-competitive energy futures can reveal trade-offs that remain hidden when focusing on a single optimum. Price~et~al.~\cite{price2017} extended this approach with novel formulations to explore near-cost-optimal configurations under alternative assumptions. Pedersen~et~al.~\cite{pedersen2021} applied MGA and related techniques to highly renewable systems, showing that diverse technology portfolios can achieve similar costs while differing in spatial and technological composition. In the context of hydrogen production, explicit applications of MGA remain limited, but recent scenario ensembles highlight its relevance. For instance, Sinha~et~al.~\cite{sinha2024} showed that hydrogen deployment diverges significantly across near-optimal decarbonization pathways for the U.S. energy system, ranging from electrolytic hydrogen to fossil-based hydrogen with carbon capture, with sectoral uptake varying widely.

Despite extensive techno-economic assessments of hydrogen import pathways, most studies remain narrowly focused on deterministic cost optimizations within techno-centered scenarios. Structural uncertainties, arising from decision-making, institutional processes, and stakeholder preferences, have received little attention, even though they can determine whether projects are viable in practice. This gap limits the ability of existing analyses to inform robust strategies for hydrogen production and transportation pathways. To address this gap, we examine how non-quantifiable factors influence the design of hydrogen import pathways. We focus on preferences that shape system configurations, including the deployment of the main technologies (PV, wind turbines, batteries, electrolyzers, and hydrogen storage) and the selection of energy carriers (hydrogen, methane, methanol, or ammonia). To capture flexibility in decision-making, we apply a Modeling to Generate Alternatives (MGA) approach~\cite{pedersen2021, LOMBARDI20202185}, which explores a wide set of diverse, near-optimal solutions within a reasonable cost slack, to an energy system optimization model for hydrogen production and transportation developed using PyPSA. The method used in this work is the Modeling All Alternatives (MAA) method~\cite{pedersen2021}, which aims to define the entire near-optimal space by constructing the convex hull that encloses it. Because the near-optimal solution space can be vast and difficult to interpret, we complement MGA with interpretable machine learning to distill these numerous results into actionable insights and to assess the impact of key decisions (e.g. excluding storage) and their consequences on alternative designs within the near-optimum cost range.

Overall, the study offers decision-makers practical guidance when unexpected location-specific, institutional, or stakeholder constraints arise regarding the deployment of renewable generation, storage capacities, or carrier options.

In this paper, we first present the hydrogen importation optimization model, the MGA approach, and the interpretable machine learning methods (\autoref{sec:methods}). The results (\autoref{sec:results}) begin with a description of the optimized solutions for the different pathways (carriers and transportation methods), along with a cost breakdown to highlight the main cost drivers for each. Next, we present the near-optimum space, demonstrating the wide variety of designs that remain within a near-optimum cost range for each pathway. Finally, we present the results from the interpretable machine learning analysis, offering key insights into the impact of decisions, such as the absence of certain technologies, on the final design configuration. We also identify clusters of solutions that point to a limited set of design archetypes that can be pursued. \autoref{sec:discussion} provides a discussion of the results and the main limitations of the study. \autoref{sec:conclusion} concludes this paper with the main takeaways.

\section{Methods}
\label{sec:methods}
The methods section begins by outlining the energy system optimization model used for the techno-economic optimization of the production and transportation pathways for each carrier: hydrogen, methane, methanol, and ammonia (\autoref{sec:trace}). \autoref{sec:case_study} then introduces the case study implemented in the model, focusing on a renewable-rich production site in Morocco and a high-demand, densely populated import location in Belgium. Next, \autoref{sec:meth_near_opt} presents the MGA approach, which generates diverse near-optimal solutions for the different pathways. This provides a set of maximally distinct alternatives within an acceptable cost range for producing and transporting hydrogen and other carriers, addressing non-quantifiable uncertainties in decision-making. Finally, \autoref{sec:meth_dt} details the interpretable machine learning method used to translate the numerous near-optimal alternatives into clear design guidelines.

\subsection{Energy system optimization model for hydrogen production and transportation}
\label{sec:trace}

The hydrogen supply chain in this study is modeled based on the TRACE model~\cite{hampp2023import}, a linear programming model developed within the PyPSA framework~\cite{PyPSA} to evaluate international energy supply chains for hydrogen and other electrofuels. TRACE determines the cost-optimal investment and operation of renewable generation, conversion, storage, and transport infrastructure to deliver a fixed energy demand.  The model represents solar and wind generation at hourly resolution and applies a greenfield assumption, where all infrastructure is newly built. It includes desalination, direct air capture (DAC), and air separation units (ASU) to supply water, carbon dioxide, and nitrogen for fuel synthesis. Buffer storage is modeled to satisfy must-run constraints in chemical synthesis plants, and losses are considered in processes such as liquefaction, compression, and boil-off. The supply chain ends at the import terminal: downstream distribution, end-use and the use of by-products are not represented.

The objective function minimizes the total annualized system cost, which includes:
\begin{itemize}
    \item Capital expenditures for generation, conversion, storage, and transport technologies ($C_\mathrm{i}$), annualized using an annuity factor based on the weighted average cost of capital (WACC) and the technical lifetime:
    \[
    A_i = \frac{(1 + r)^{n_\mathrm{i}} \cdot r}{(1 + r)^{n_\mathrm{i}} - 1},
    \]
    where $r$ is the interest rate, $\mathrm{i}$ denotes the component index and $n_\mathrm{i}$ its lifetime.
    \item Fixed operation and maintenance (O\&M) costs ($O_\mathrm{i}$).
\end{itemize}

The annualized cost for each component is therefore:
\[
c_\mathrm{i} = C_\mathrm{i} \cdot A_\mathrm{i} + O_\mathrm{i},
\]
and the optimization problem is formulated as:
\[
\underset{x_\mathrm{i}}{\text{minimize}} \sum_{\mathrm{i}} c_\mathrm{i} \cdot x_\mathrm{i}.
\]
where $x_\mathrm{i}$ denotes the capacity of each component to be optimized.

\begin{figure}[h!]
    \centering
    \includegraphics[width=0.9\linewidth]{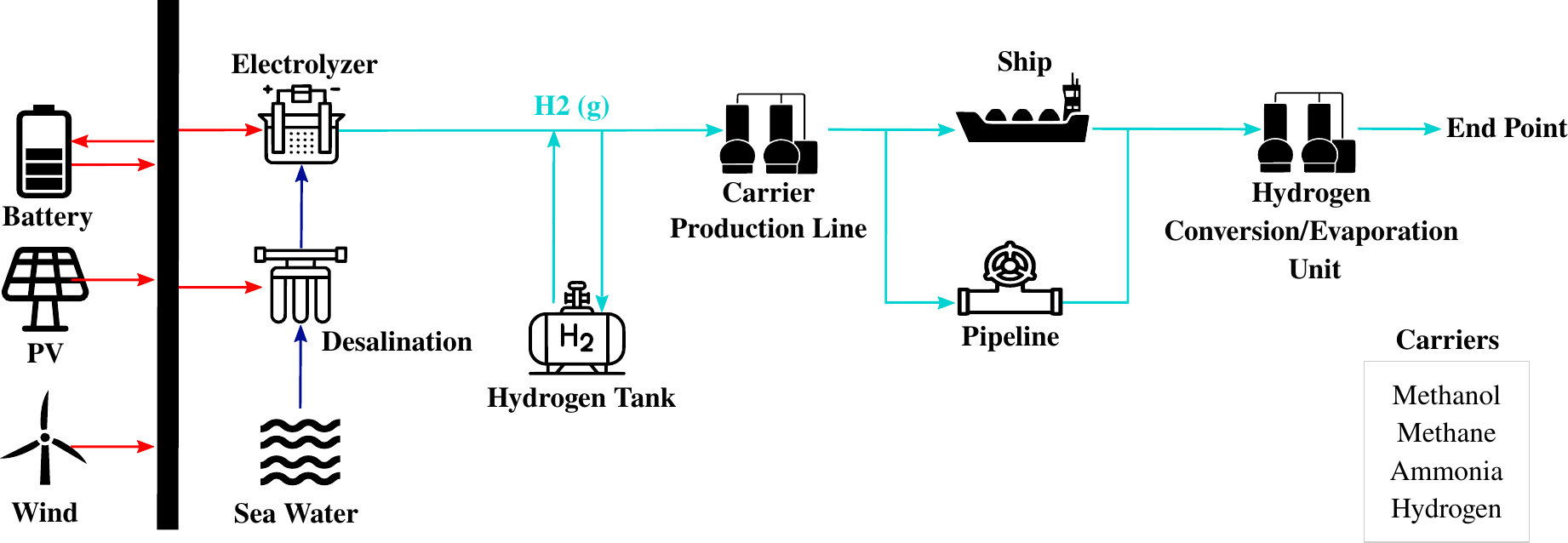}
    \caption{Overview of the hydrogen import supply chain. Renewable electricity from solar PV and wind turbines is used to produce hydrogen via PEM electrolysis, with seawater desalination providing the required feedwater. Battery storage balances the short-term variability of renewable generation, while hydrogen storage buffers supply before conversion. The produced hydrogen can then enter one of four carrier-specific synthesis pathways: (i) compression or liquefaction for direct hydrogen transport, (ii) catalytic methanation with captured $\text{CO}_2$ to produce synthetic methane, (iii) combination with nitrogen from an air separation unit in a Haber-Bosch process to produce ammonia, or (iv) catalytic synthesis with $\text{CO}_2$ to produce methanol. The resulting carriers are transported either by ship or pipeline to the import terminal. At the import side, carriers go through reconversion: regasification for liquid hydrogen, reforming for methane and methanol, and cracking for ammonia, so that all pathways ultimately deliver gaseous hydrogen as a uniform end product.}
    \label{fig:h2_sc}
\end{figure}

The supply chain is structured as follows. Renewable electricity is generated from solar and wind sources (PV panels and wind turbines), while storage technologies (batteries and hydrogen storage) are included to mitigate intermittency, as the system is modeled as an islanded configuration with no connection to the power grid. Additional storage for liquid carriers is installed at both the export and import ports to act as buffers before and after shipping. Hydrogen is produced through PEM electrolysis, coupled with a desalination unit that supplies pure water from seawater. From this point, our analysis considers four possible hydrogen carriers: hydrogen, methane, ammonia, and methanol, and two transportation options: pipeline and shipping. Hydrogen can either be transported directly in gaseous form by pipeline or liquefied and shipped, with boil-off losses accounted for during shipping. Methane is produced via catalytic methanation of hydrogen with carbon dioxide captured from the air from DAC, and is either transported by pipeline in gaseous form or liquefied for shipping. Ammonia is synthesized through the Haber–Bosch process, combining hydrogen with nitrogen from an ASU, transported as liquid either by pipeline or ship. Methanol is produced from hydrogen and carbon dioxide through catalytic synthesis and can likewise be transported by pipeline or ship. At the import terminal, all carriers are converted back to gaseous hydrogen to provide a uniform end product: liquid hydrogen is evaporated, ammonia is cracked, and methane and methanol are reformed. The potential use of by-products from these conversion processes is not accounted for in the model. All techno-economic parameters, including process efficiencies, energy demands, and cost assumptions, follow the values reported in Hampp~et~al.~\cite{hampp2023import}. A schematic overview of the modeled supply chain is shown in \autoref{fig:h2_sc}.

\subsection{Case study}
\label{sec:case_study}

\begin{figure}[h!]
    \centering
    \includegraphics[width=0.8\linewidth]{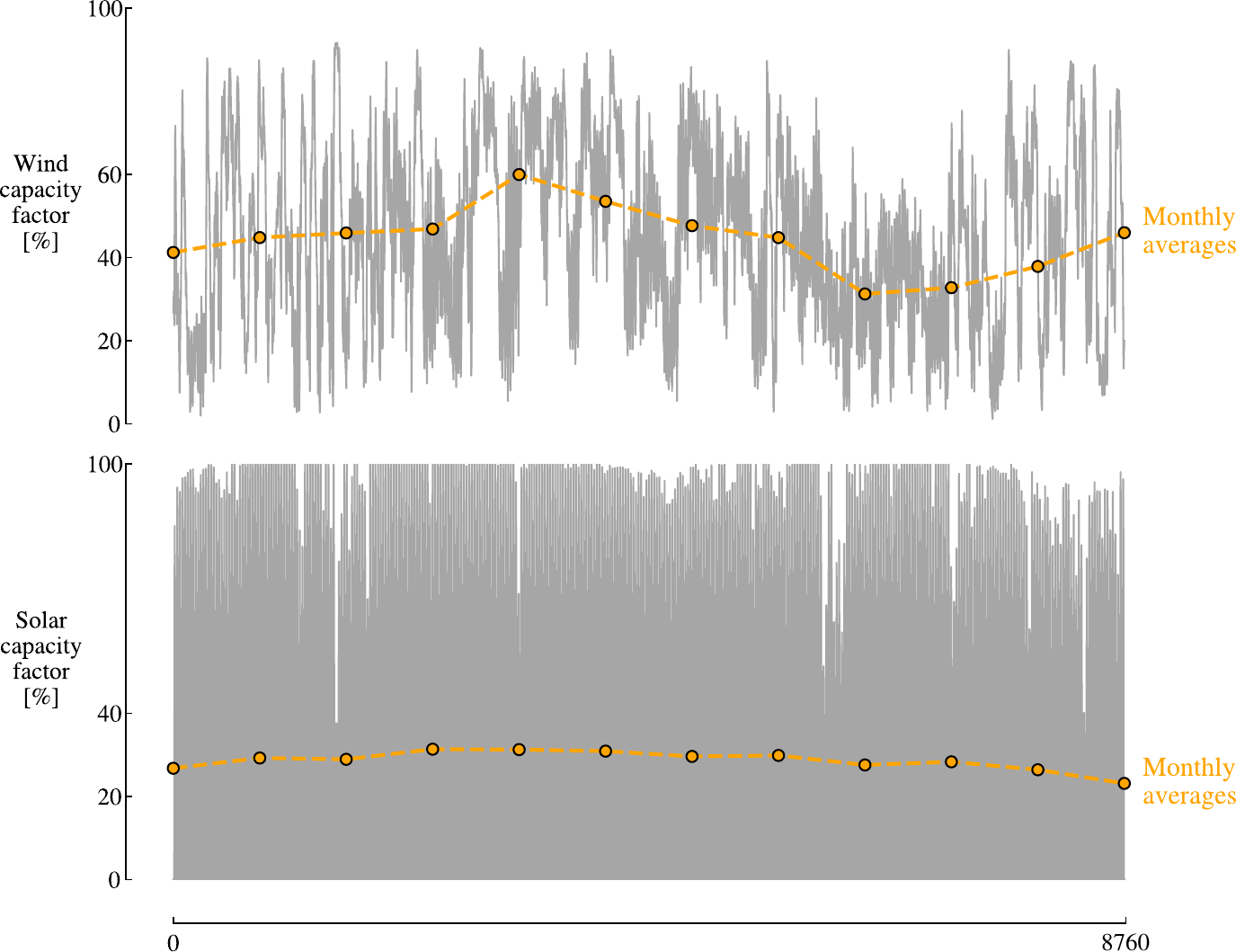}
    \caption{Renewable potential of Morocco, showing high wind and solar resources with average capacity factors of 44\% and 29\%, respectively. The monthly average of solar energy remains stable year-round, whereas wind energy is more variable, with the highest average potential during spring.}
    \label{fig:renewables_potential}
\end{figure}

The case study considers an export-import corridor between Morocco and Belgium. Morocco is selected as the production site due to its abundant solar and wind resources, its established role as a front-runner in renewable energy deployment, and its strategic geographic position at the interface between Africa and Europe~\cite{morocco_choice}. The renewable potential of Morocco is illustrated in \autoref{fig:renewables_potential}. The average capacity factor is 44\% for wind and 29\% for solar. While the monthly average solar energy remains consistent throughout the year, the monthly average wind energy exhibits seasonal variability, with higher output during the spring months.

Belgium is chosen as the import destination, with an annual hydrogen import target of \SI{20}{TWh_{H_2}} (\SI{0.6}{Mt_{H_2}}), consistent with the ambition set in the Belgian federal hydrogen strategy~\cite{import_cap}. The corridor represents a mid-distance trade route of 2517~km.

Overall, this study evaluates eight potential hydrogen import pathways from Morocco to Belgium: four carriers and two transport modes, each optimized independently, resulting in eight distinct optimal designs.
\subsection{Modeling to Generate Alternatives}
\label{sec:meth_near_opt}
Energy system optimization models typically return a single least-cost solution. While this benchmark is useful, it is unlikely to be realized in practice. Technology costs are uncertain, performance differs across contexts, and political or social preferences may shift designs away from the mathematical optimum. It is therefore more informative to examine not only the strict cost-optimal configuration but also the broader set of near-optimal designs: solutions that are only slightly more expensive but structurally different. This helps identify which technologies are required across all cost-competitive designs and which are flexible, allowing trade-offs without large cost penalties.  

The Modeling to Generate Alternatives (MGA) approach provides a structured way to achieve this. Following Pedersen~et~al.~\cite{pedersen2021}, we adopt the Modeling to Generate All Alternatives (MAA) method, which systematically maps the near-optimal solution space. Formally, let $f(x)$ denote the objective function and $x^*$ the optimal solution. The near-optimal space $W$ is defined as:  
\begin{equation}
W = \{\, x \in X \; | \; f(x) \leq f(x^*) (1 + \varepsilon) \,\},
\end{equation}
where $X$ is the feasible solution space and $\varepsilon$ is the slack parameter. This ensures that only solutions within a specified tolerance of the optimum are explored. 
In this study, \(f(x)\) corresponds to the total annualized system cost, and \(\varepsilon\) is set to \(10\%\) for each pathway individually, meaning that all solutions within \(10\%\) of the least-cost design of that pathway are considered near-optimal.

The algorithm then proceeds in two phases. The first phase defines the boundaries of the near-optimal space by constructing the convex hull~\cite{2013ascl.soft04016B}, which is the smallest convex polyhedron enclosing all near-optimal solutions. To generate the points that define this convex hull, the objective function is modified from minimizing the total system cost to minimizing a weighted sum of the MGA variables of interest: PV, wind, electrolyzer, battery, and hydrogen storage tank capacities, while enforcing the MGA cost constraint:
\begin{equation}
\begin{aligned}
\underset{x_\mathrm{i},\, \mathrm{i} \in V_\mathrm{MGA}}{\text{minimize}} \quad & \sum_{\mathrm{i}} w_\mathrm{i} \cdot x_\mathrm{i} \\
\text{s.t.} \quad & f(x) \leq f(x^*) (1 + \varepsilon).
\end{aligned}
\end{equation}

The process begins by maximizing and minimizing each of the MGA variables and constructing a convex hull around the resulting solutions using the Quickhull algorithm. At each iteration, a set of $m$ weight vectors $\{w_k\}_{1\leq k \leq m}$ is generated such that they are normal to the faces of the convex hull obtained in the previous iteration. Using these weights, new points are explored, and a new convex hull is then constructed that encompasses both previously identified and newly identified solutions. This procedure is repeated, and the convex hull is iteratively updated until its volume converges.

Once the boundaries are mapped, the second phase samples solutions from the interior of the polyhedron. Because the whole problem is convex, any point inside the convex hull is also feasible and near-optimal. The polyhedron is divided into simplexes using the Qhull software~\cite{2013ascl.soft04016B}, and random samples are drawn within these simplexes by computing a weighted sum of the vertices defining each simplex. Pedersen~et~al.~\cite{pedersen2021} used the Bayesian bootstrap method to generate the weights, ensuring an even spread of points across the region. The outcome is a large set of alternative system designs that are distinct in structure, but all remain cost-competitive.  

This method avoids the limitations of other MGA approaches, which often produce only a handful of alternatives without systematically exploring the entire near-optimal space, thereby risking the omission of important solutions. In contrast, the MAA method provides a comprehensive coverage of the near-optimal region while efficiently generating a large number of alternative solutions within seconds.

The MGA procedure is applied separately to each of the eight pathways described in \autoref{sec:case_study}, generating a diverse set of near-optimal designs for each pathway. This enables us to explore design trade-offs, such as between wind and PV, and to identify which technology choices are interdependent.

\subsection{Interpretable machine learning}
\label{sec:meth_dt}


The MGA approach (\autoref{sec:meth_near_opt}) generates a large set (in the order of $10^{6}$ with MAA) of near-optimal solutions with only a slight increase in cost. While this set provides a broad view of feasible and acceptable designs, the large number of solutions can make interpreting decisions and extracting meaningful insights a challenging task for decision-makers. To address this, we adopt the interpretable machine learning framework of \cite{streamliningenergytransitionscenarios}, which translates a high-dimensional solution set into a small number of clear and interpretable decision rules.  

The method begins by applying clustering algorithms to group the solutions. Each solution is assigned a cluster label, forming sets of configurations with similar characteristics. In our analysis, the decision variables consist of categorical variables, which are the binary indicators specifying the chosen transport carrier, and continuous variables representing the installed capacities of PV, wind turbines, electrolyzers, batteries, and hydrogen storage. When only continuous variables are involved, we use the \(k\)-means algorithm, whereas when both binary and continuous variables are present, we use \(k\)-prototypes clustering~\cite{huang1998}. This method extends \(k\)-means by combining squared Euclidean distances for continuous variables with a simple matching dissimilarity for categorical variables. The overall dissimilarity between design \(x_i\) and cluster centroid \(c_j\) is defined as:  
\begin{equation}
d(x_i, c_j) = \sum_{l \in \mathcal{C}} \bigl(x_{i l} - c_{j l}\bigr)^2 \;+\; \gamma \sum_{m \in \mathcal{D}} \delta(x_{i m}, c_{j m}),
\end{equation}
where \(x_i\) is the normalized vector of decision variables for design \(i\), and \(c_j\) is the centroid of cluster \(j\). The index sets \(\mathcal{C}\) and \(\mathcal{D}\) refer to continuous and categorical variables, respectively. For a continuous variable \(l \in \mathcal{C}\), the squared difference \((x_{i l} - c_{j l})^2\) measures the distance in that dimension. For a categorical variable \(m \in \mathcal{D}\), the function \(\delta(x_{i m}, c_{j m})\) equals zero if design \(i\) and centroid \(j\) share the same category, and one otherwise. The parameter \(\gamma > 0\) controls the relative weight of categorical mismatches compared to continuous differences, ensuring that both binary carrier choices and continuous capacity decisions contribute to the clustering outcome. Low \(\gamma\) values give more influence to continuous variables, whereas high values make categorical features dominant. In our case, \(\gamma\) was set to \(0.02\). The automatic value obtained using the method proposed in~\cite{Huang1997CLUSTERINGLD} is \(\gamma = 0.07\). After testing other values close to this value, we selected \(0.02\) because it produced clusters that were easier to interpret in later analyses.

In the second step, a decision tree (Classification and Regression Tree - CART) is trained~\cite{breiman1984cart}, with the MGA decision variables as features and the cluster labels as targets. The decision tree identifies splits in the feature space by defining thresholds on feature values, ultimately mapping designs to their respective clusters. This yields a compact set of hierarchical rules that explain how different clusters arise.  

The final step involves reassigning designs based on the decision tree's predictions. While most designs remain in their original clusters, those classified differently by the decision tree are reassigned accordingly. After this step, the clusters are defined by the splits and thresholds determined by the decision tree, providing a more interpretable and structured representation of the near-optimal space.  

The selection of the number of clusters and the depth of the decision tree is based on a balance between accuracy and interpretability, and is detailed in the Supplementary Information.

We first apply this procedure using \(k\)-prototypes clustering on the full set of solutions generated by the MGA approach across the three non-hydrogen shipping pathways (corresponding to ammonia, methane and methanol). This allows the clustering to capture both the design characteristics of each solution and the associated carrier choice. Similar patterns were observed for the pipeline case, therefore, we present only the shipping results here. The pipeline results can be found in the Supplementary Information. To further examine the structure of the near-optimal space, we then use \(k\)-means clustering separately to each individual pathway. This second step captures within-pathway trade-offs, such as different balances between renewable generation and storage. Taken together, these two stages provide both a cross-pathway view of the solution space and a detailed understanding of the trade-offs within each pathway.

\section{Results}

\label{sec:results}
In this section, we first discuss the cost-optimal configuration for each import pathway, followed by a detailed cost breakdown (\autoref{subsec:optimal_solutions}). We then showcase the near-optimal solution space (\autoref{subsec:near_optimum_space}), and finally streamline the large set of designs identified through this exploration using interpretable machine learning (\autoref{subsec:streamlining_decisions}).

\subsection{Cost-optimal hydrogen import pathways}
\label{subsec:optimal_solutions}

\begin{figure}[h!]
    \centering
    \includegraphics[width=0.5\linewidth]{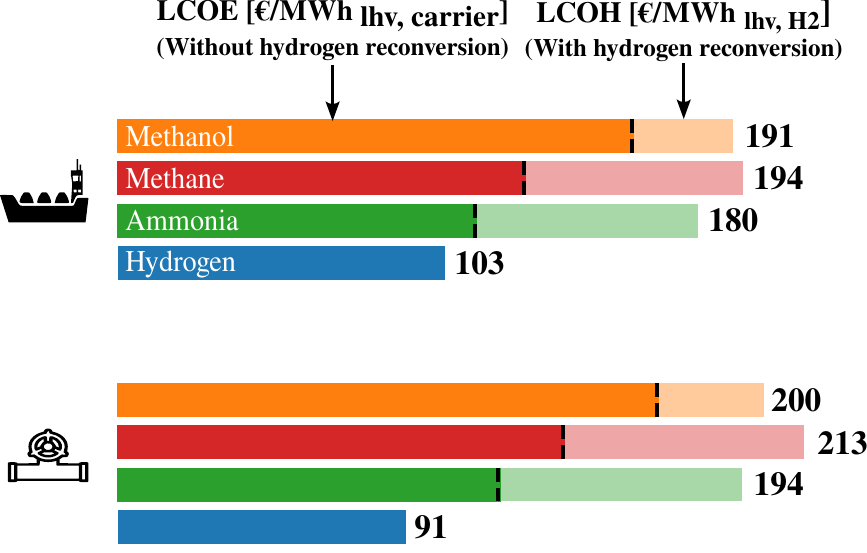}
    \caption{Hydrogen pathways (shipping and pipelines) are associated with the lowest LCOH, while non-hydrogen carrier pathways exhibit nearly twice the cost. Among non-hydrogen options, ammonia has the lowest LCOH, while methane has the highest. Excluding the reconversion step for non-hydrogen carriers significantly reduces the LCOE difference, bringing them closer to hydrogen and making methane cheaper than methanol. This contrasts with the case including the conversion step, where methanol performs better as a hydrogen carrier, whereas methane appears more suitable as a direct energy source.}
    \label{fig:opt_lcoe}
\end{figure}

As a first step, we generate the cost-optimal solutions for each hydrogen import pathway. Hydrogen carrier pathways exhibit the lowest LCOH, while non-hydrogen carriers: ammonia, methane, and methanol, show roughly twice the cost (\autoref{fig:opt_lcoe}).
In particular, hydrogen transported through pipelines achieves the lowest LCOH (91~\EUR{}/MWh), while non-hydrogen carriers range from 180~\EUR{}/MWh for ammonia (shipping transport) to 213~\EUR{}/MWh for methane (pipeline transport), representing the upper end of the LCOH range. This outcome is expected because non-hydrogen pathways require additional synthesis and reconversion stages, such as ammonia synthesis, methanation, or methanolization, followed by cracking or reforming at the import terminal. These extra units increase capital costs and introduce conversion losses, leading to higher energy consumption per unit of delivered hydrogen. 

These results are consistent with the LCOH reported by Hampp~et~al.~\cite{hampp2023import}, ranging from 3~\EUR{}/kg\(_\mathrm{H_2}\) ($\approx$	 100~\EUR{}/MWh\(_\mathrm{H_2}\)) for hydrogen transported via pipelines, to 3.4~\EUR{}/kg\(_\mathrm{H_2}\) ($\approx$	 113~\EUR{}/MWh\(_\mathrm{H_2}\)) for hydrogen via shipping. For non-hydrogen carriers, they report 6~\EUR{}/kg\(_\mathrm{H_2}\) ($\approx$200~\EUR{}/MWh\(_\mathrm{H_2}\)) for ammonia via shipping, 6.2~\EUR{}/kg\(_\mathrm{H_2}\) ($\approx$206~\EUR{}/MWh\(_\mathrm{H_2}\)) for methanol via shipping, and 7~\EUR{}/kg\(_\mathrm{H_2}\) ($\approx$233~\EUR{}/MWh\(_\mathrm{H_2}\)) for methane via pipelines.

Despite the significant gap in LCOH between hydrogen and non-hydrogen carriers, these carriers may still be considered attractive because existing infrastructure and possible integration with other industrial systems (such as ammonia and methane networks) can reduce overall costs~\cite{CUI202315737}, even though this is not captured in our model. Moreover, decades of operational experience with ammonia and methane handling may make them more appealing despite their higher costs, especially when considering the safety risks associated with hydrogen~\cite{CUI202315737}.

On the transport side, shipping is generally more cost-effective than pipelines, except for hydrogen, where pipeline delivery avoids liquefaction losses and thus achieves a lower LCOH. Among the non-hydrogen options, ammonia remains the most competitive due to mature synthesis and cracking processes, followed by methanol, and finally methane.

When the conversion back to hydrogen is excluded, the levelized cost of energy (LCOE,~\euro{}/MWh$_\text{LHV, carrier}$) shows a smaller cost gap between hydrogen and the other chemicals. In this case, ammonia reaches costs comparable to hydrogen. And although methane has a higher LCOH than methanol, its LCOE is actually lower, suggesting that methane may be better suited for direct energy use while methanol may be more appropriate as a hydrogen carrier. This outcome could change if existing fossil gas infrastructure were taken into account, as this study adopts a greenfield approach where no prior infrastructure is considered. In such a case, methane could become more competitive. Evaluating carriers based on their LCOE rather than their reconverted hydrogen cost would involve fundamentally different system boundaries, addressing chemical-specific uses, exergy content, and sectoral demand (e.g., ammonia as fertilizer feedstock, methane for combustion, or hydrogen for industrial reduction). Such cross-sectoral assessments fall outside the scope of this study, which is strictly focused on pathways that deliver hydrogen as the final product after reconversion. 

\begin{figure}[h!]
    \centering
    \includegraphics[width=0.8\linewidth]{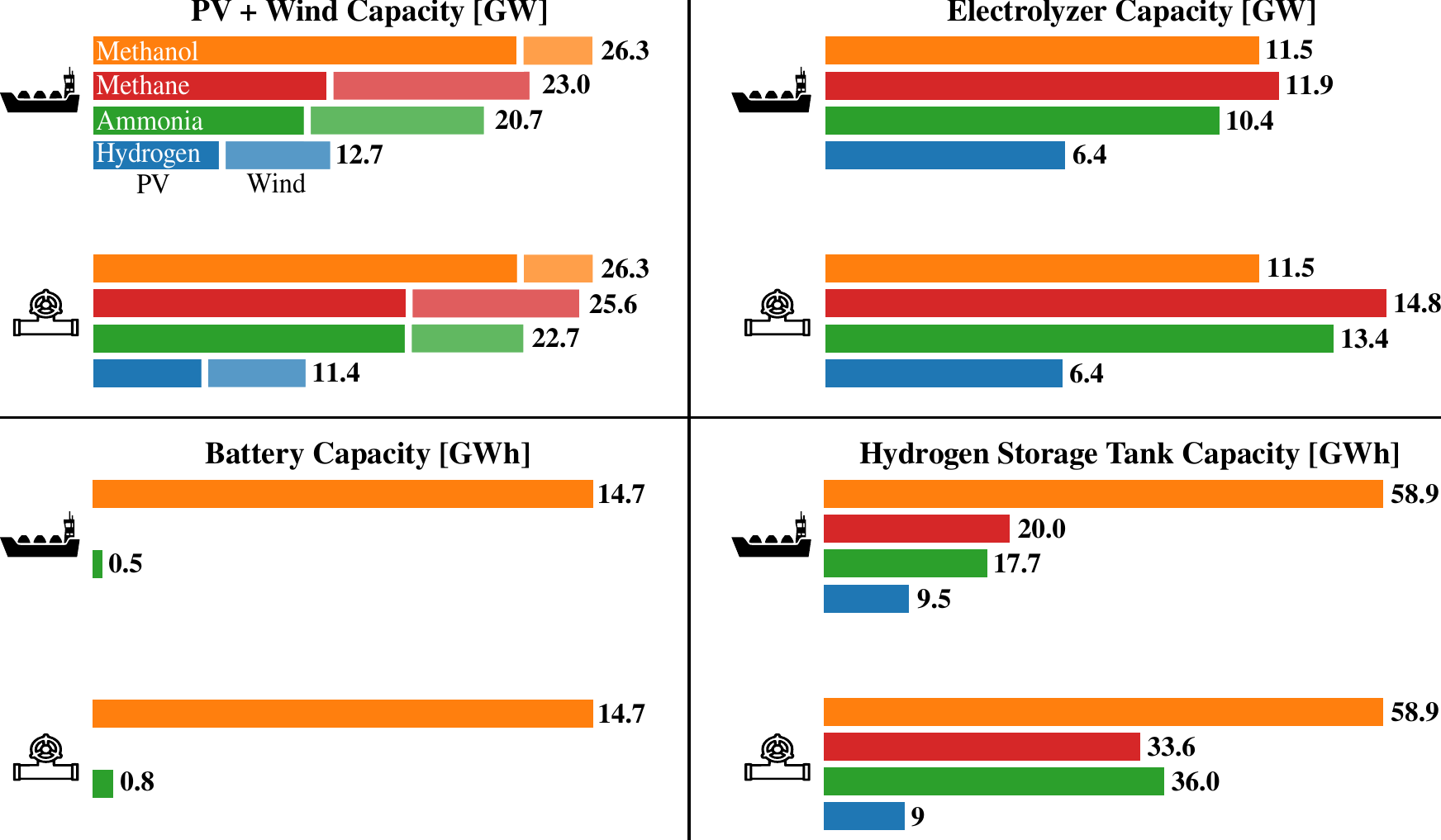}
    \caption{Installed capacities for the optimized pathways, showing that the hydrogen pathway requires the lowest overall capacities, while methanol has the largest energy storage and electricity generation capacities. Methane exhibits the highest electrolyzer capacity. The high storage and generation requirements for methanol come from the minimum power/hydrogen intake flow constraint of the methanolization unit.}
    \label{fig:opt_caps}
\end{figure}

Moving on to the optimal configuration of each pathway, we observe that, consistent with the cost results, hydrogen carrier pathways are the least investment-demanding across all technologies (\autoref{fig:opt_caps}).

\noindent Among the non-hydrogen carriers, ammonia requires the lowest PV and wind capacities: 20.7~GW of total PV and wind, followed by methane with 23.0~GW of total PV and wind, and then methanol, which requires the largest generation capacities at 26.3~GW of total PV and wind with shipping as transportation option. Regarding the electrolyzer, methane surpasses methanol in the shipping pathway, with 11.9~GW compared to 11.5~GW, while in the pipeline pathway both ammonia (13.4~GW) and methane (14.8~GW) exceed methanol (11.5~GW). For hydrogen storage, methane and ammonia exhibit similar capacities of 20.0~GWh and 17.7~GWh in the shipping pathway, respectively, whereas methanol requires significantly higher storage at 58.9~GWh. A similar pattern is observed for batteries: hydrogen, ammonia, and methane pathways have near-zero battery capacities (below 1~GWh), while methanol reaches 14.7~GWh. This high electricity production and storage requirement for methanol is due to the fact that the methanol synthesis unit needs to operate continuously at a minimum flow.

The differences between hydrogen and non-hydrogen pathways are largely explained by the conversion losses at the destination, which for non-hydrogen carriers significantly outweigh the additional energy required for hydrogen liquefaction in the liquid hydrogen (shipping) pathway. This is reflected in the energy consumption of the different hydrogen production and transportation pathways (defined as the ratio of total generated renewable electricity to the delivered hydrogen energy) shown in \autoref{tab:round_trip}. For hydrogen, the energy consumption ranges from \SI{1.64}{MWh_{elec}/MWh_{H_2}} to \SI{1.81}{MWh_{elec}/MWh_{H_2}}, whereas for the other carriers it ranges from \SI{2.75}{MWh_{elec}/MWh_{H_2}} to \SI{3.52}{MWh_{elec}/MWh_{H_2}}.

\begin{table}[h!]
\centering
\caption{Energy consumption for each pathway [\SI{}{MWh_{elec}}/\SI{}{MWh_{LHV, H_2}}]}
\label{tab:round_trip}
\begin{tabular}{lcccc}
\toprule
 & \textbf{Hydrogen} & \textbf{Ammonia} & \textbf{Methane} & \textbf{Methanol} \\
\midrule
\textbf{Shipping} & 1.81 & 2.97 & 3.33 & 2.76 \\
\textbf{Pipeline} & 1.64 & 2.98 & 3.52 & 2.75 \\
\bottomrule
\end{tabular}
\end{table}

\begin{figure}[h!]
    \centering
    \includegraphics[width=1\linewidth]{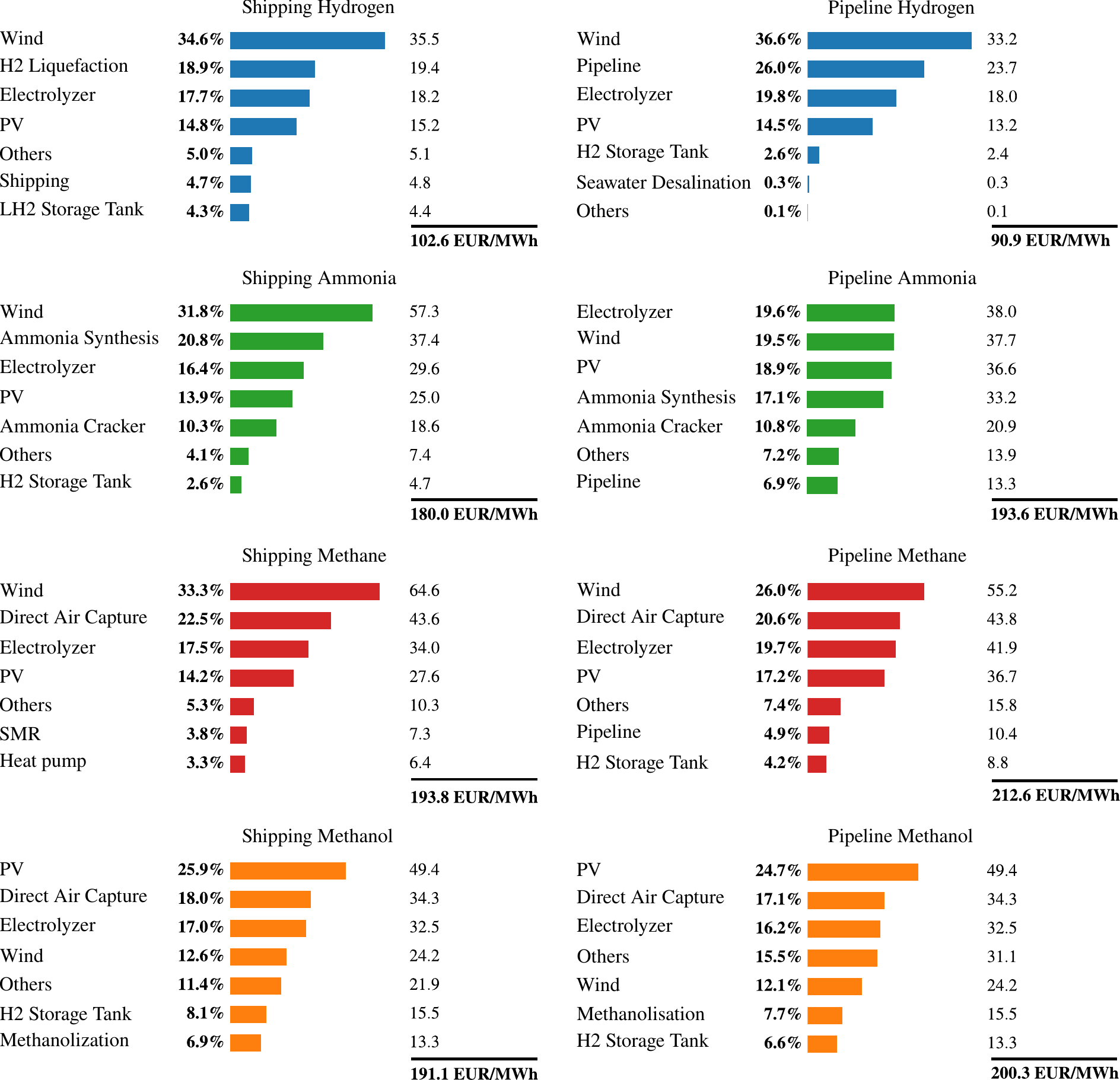}
    \caption{Cost breakdown of the optimized pathways, showing that wind dominates the costs for hydrogen, ammonia, and methane, while PV is the main cost driver for methanol. Non-hydrogen carriers face higher overall costs due to additional conversion units (e.g., ammonia synthesis, direct air capture), which add to the already higher costs of shared technologies such as wind and electrolysis.}
    \label{fig:cost_breakdown}
\end{figure}

The cost breakdown shows that the main cost driver is different for the different hydrogen production and transportation pathways (\autoref{fig:cost_breakdown}). 
For hydrogen, ammonia, and methane, wind installations account for the largest share of total costs, while for methanol, PV installations dominate, showing the critical role of solar generation in methanol production. The electrolyzer typically ranks third, except in the ammonia pipeline pathway, where it becomes the main contributor, followed by wind and PV. The second-largest contributors are carrier-specific technologies: hydrogen liquefaction for hydrogen shipping, direct air capture for carbon-based carriers, and ammonia synthesis for ammonia. Other major contributors include ammonia cracking for ammonia, steam reforming and heat pumps for methane, and methanol synthesis for methanol. These conversion-related technologies substantially raise the LCOH, exceeding the additional costs of hydrogen-specific processes such as liquefaction and liquid hydrogen storage. Furthermore, even shared technologies like wind and electrolysis cost more in non-hydrogen pathways than in the hydrogen pathway, consistent with the higher capacities installed in these pathways.

\subsection{Near-optimal solution space}
\label{subsec:near_optimum_space}

\begin{figure}[h!]
    \centering
    \includegraphics[width=0.8\linewidth]{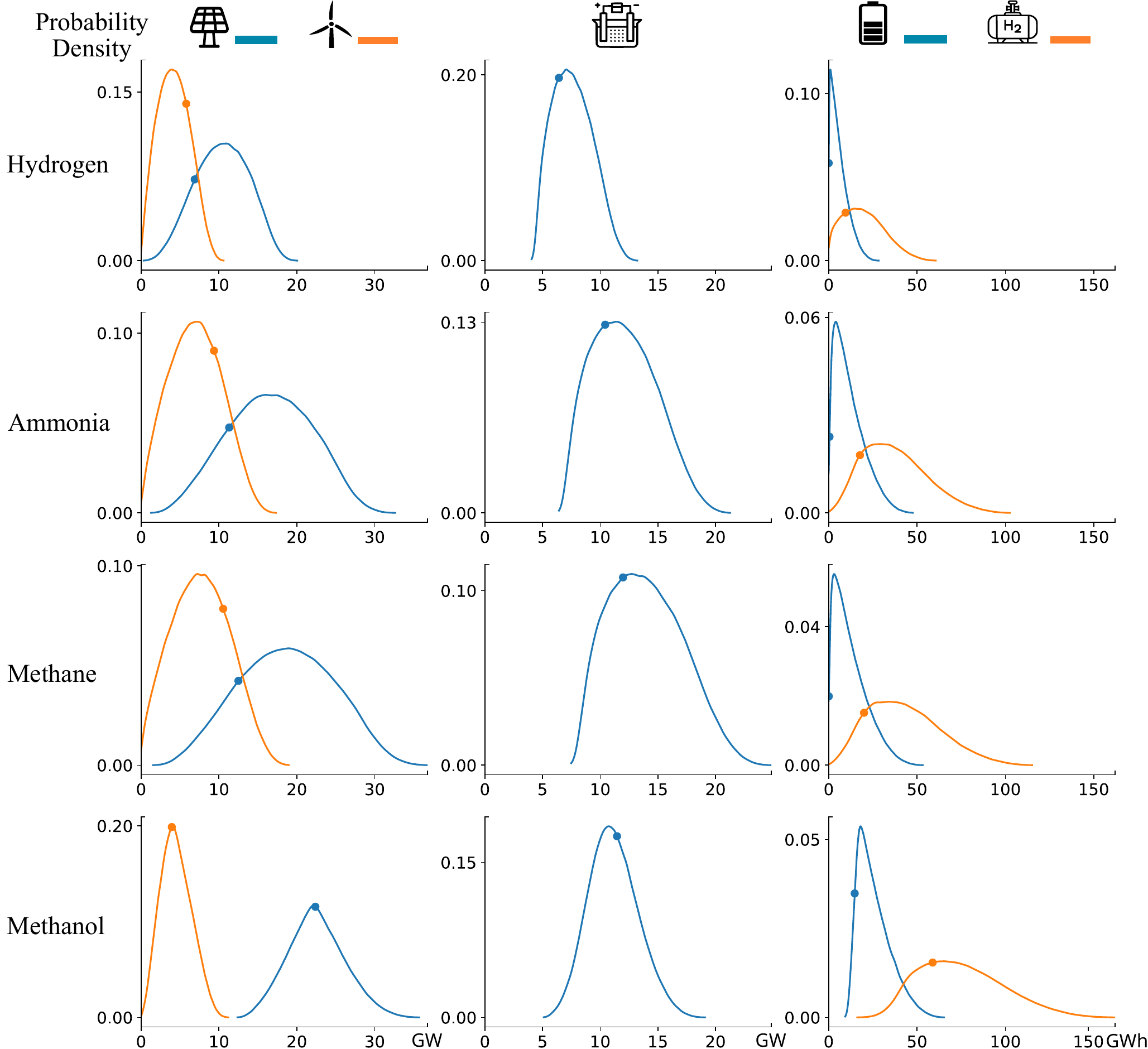}
    \caption{Probability density plots of the near-optimal solution space for the shipping pathways. The distributions illustrate the wide range of feasible capacities that remain within 10\% of the minimum LCOH. The dot indicates the cost-optimal configuration. Across carriers, the near-optimal designs exhibit great diversity: PV, wind, and electrolyzer capacities vary widely, while battery storage appears in near-optimal configurations despite being absent in the optimal ones. Overall, there are no 'must-have' technologies, as most technologies (except the electrolyzer) can be entirely excluded for most carriers.}
    \label{fig:near_opt_space}
\end{figure}

Following the application of the MGA to the energy system optimization model for each hydrogen production and transportation pathway, the near-optimal space for technology capacities reveals great diversity (\autoref{fig:near_opt_space}). Wind capacity, for instance, ranges from zero up to about 10~GW for the hydrogen and methanol pathways, and reaches as high as 20~GW for methane and ammonia. PV capacity shows an even broader range, extending from zero to roughly 20~GW for hydrogen, up to 35~GW for ammonia, and around 40~GWp for methane and methanol. However, for methanol, PV capacity does not reach zero, but instead starts at around 12~GWp. Electrolyzer capacity also exhibits great flexibility: while it cannot reach zero, as hydrogen production is essential, it varies between roughly 5~GW and 25~GW depending on the carrier. Regarding storage capacities, battery storage, which is absent in the cost-optimal configurations for hydrogen, ammonia, and methane, emerges in near-optimal designs with capacities reaching up to 60~GWh. Hydrogen storage, on the other hand, ranges from zero to 50~GWh for hydrogen, 100~GWh for ammonia and methane, and, for methanol, never reaches zero, varying instead between 16~GWh and 162~GWh.

What stands out here is that both generation and storage technologies can, in theory, be entirely excluded for hydrogen, ammonia, and methane systems, while electrolyzer capacity can be reduced to roughly 60\% of its optimal value. Hence, for these three carriers, there are no 'must-have' technologies apart from the electrolyzer. Methanol systems, however, deviate from this pattern, requiring non-zero minimum capacities for both PV and hydrogen storage. It is important to note that system design cannot be approached as simple 'cherry picking'. In fact, when one technology is constrained or excluded, trade-offs arise that reshape the capacities required for others. This will be further discussed in \autoref{subsec:streamlining_decisions}.

Finally, the cost-optimal configuration (indicated by the dot in \autoref{fig:near_opt_space}) does not coincide with the mode of the near-optimal distributions. This indicates that the optimal design is not the most typical configuration within the design space. In other words, while the cost-optimal point minimizes the objective function, it represents a relatively less common combination of technology capacities. By contrast, the mode (or its neighboring region) corresponds to configurations that appear more frequently across the near-optimal solutions. Thus, focusing solely on the mathematical optimum risks overlooking the diversity of equally competitive designs that may better align with alternative technical, institutional, or stakeholder preferences.

\subsection{Turning the near-optimal solution space into actionable decisions}
\label{subsec:streamlining_decisions}

As shown in \autoref{subsec:near_optimum_space}, the near-optimal space includes a wide variety of technology combinations, where multiple system configurations achieve similar costs and many technologies are not strictly required. While this diversity highlights the flexibility of hydrogen production and transportation pathways, it also complicates interpretation, as the relationships between technologies are highly interdependent. To extract actionable insights from this large set of solutions, we apply clustering and decision trees (as detailed in \autoref{sec:meth_dt}) to identify the most influential decisions and clarify how constraining or excluding certain technologies affects the remaining system design. The analysis presented here focuses on the shipping pathways, as they capture the main design trade-offs across carriers. Similar patterns were found for the pipeline pathways, but are omitted for brevity.

We first apply the \(k\)-prototypes clustering and decision tree methods to the non-hydrogen carriers dataset to identify the main decisions across import pathways (\autoref{sec:carrier_classification}). We then perform a carrier-specific analysis using \(k\)-means clustering and decision trees to explore pathway-specific trade-offs in greater detail (\autoref{sec:carrier_specific_analysis}).

\subsubsection{Decisions across non-hydrogen import pathways}
\label{sec:carrier_classification}

\begin{figure}[h!]
    \centering
    \makebox[1\textwidth]{
        \includegraphics[width=1\linewidth]{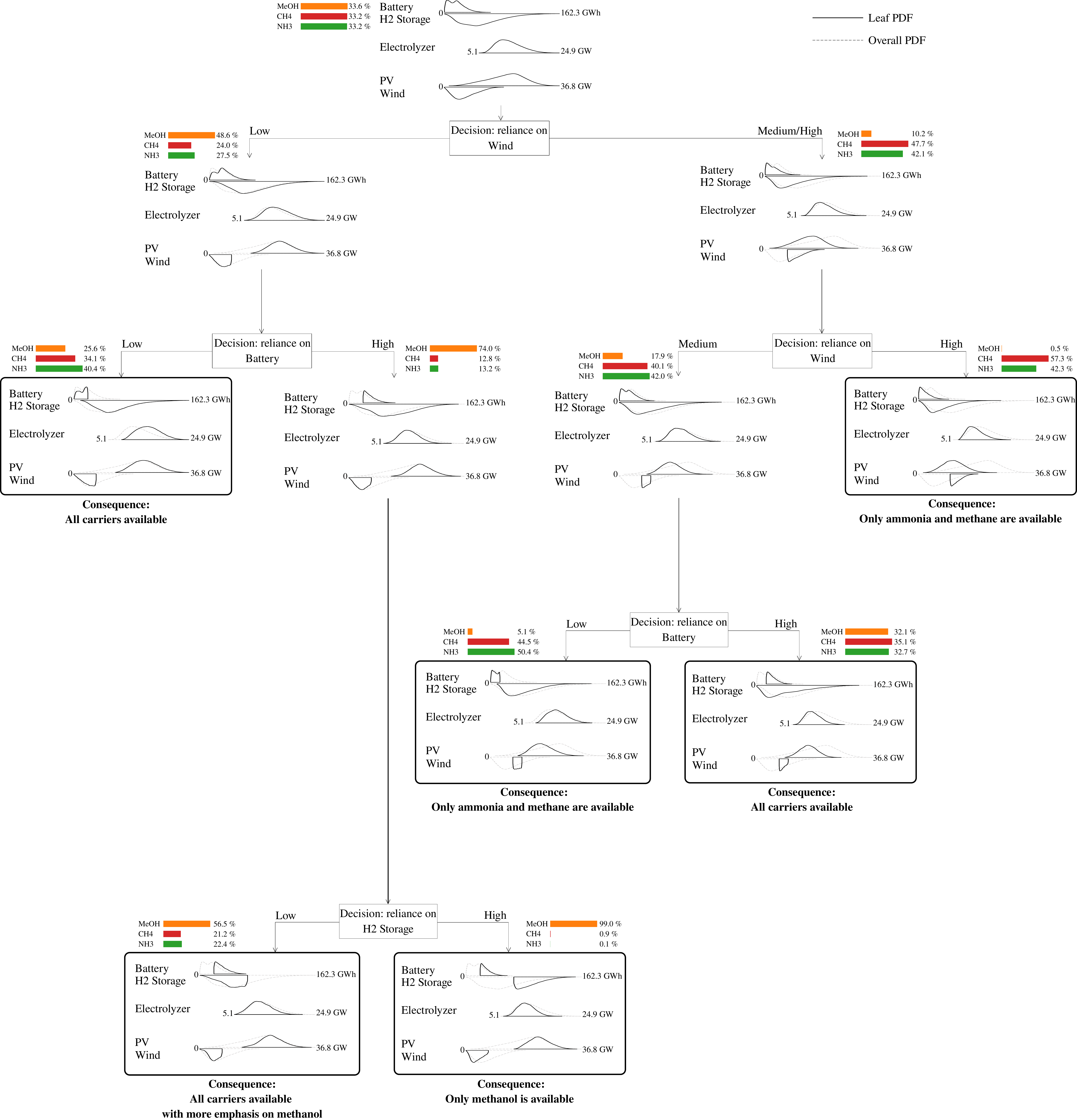}
    }
    \caption{%
    Fuel classification decision tree showing how renewable and storage capacities shape the selection of non-hydrogen-carrier import pathways within the near-optimal space. High wind capacities lead primarily to ammonia and methane pathways with low PV and storage levels, while low wind and high PV capacities promote methanol, particularly when combined with large battery and hydrogen storage. 
    At medium wind capacities, battery capacity becomes the key determinant: low battery capacity again favors ammonia and methane, whereas high battery capacity allows all three carriers to remain feasible. The tree therefore illustrates three design archetypes: wind-dominated systems using ammonia or methane, mixed configurations allowing all carriers, and PV-dominated systems relying on methanol.}
    \label{fig:fuel_classification_dt}
\end{figure}

\autoref{fig:fuel_classification_dt} illustrates how system design decisions influence the choice among non-hydrogen carriers: ammonia, methane, and methanol. At this stage, we restrict the analysis to non-hydrogen carriers because hydrogen systems differ significantly in both cost and capacity. Here, only shipping pathways are presented. The corresponding decision tree for the pipeline pathways, which follows the same trends as the shipping tree, is provided in the Supplementary Information. Each node in the decision tree splits the solution space according to a key design variable (i.e., the features): PV, wind, electrolyzer, battery, or hydrogen storage capacity.

The first and most influential decision concerns wind capacity. At the far right branch, where wind capacity is high, only ammonia and methane remain viable import pathways, whereas methanol is no longer an option. High wind generation reduces the need for large PV and storage capacities and slightly lowers electrolyzer capacity. Since methanol synthesis requires stable electricity and hydrogen inputs, and is strongly coupled with storage and PV generation, it becomes incompatible with wind-dominated configurations within the cost tolerance of 10\%. Therefore, projects emphasizing large wind farms, whether driven by resource potential, policy, or stakeholder preference, are primarily limited to ammonia or methane-based imports.

At medium wind capacities, the next decisive variable is battery storage. When battery capacity is low, the system again converges toward ammonia and methane pathways, as limited storage prevents the stable operation needed for methanol synthesis. When battery capacity is high, however, all three carriers remain feasible. The additional flexibility provided by batteries mitigates renewable variability, reopening the possibility of methanol as a carrier. In these cases, hydrogen storage can often be reduced or even omitted, since batteries smooth the power supply to the electrolyzer and synthesis units.

Moving to the left branch of the tree, where wind capacity is low and PV generation dominates, methanol re-emerges as the most frequent carrier. Under these PV-dominated conditions, low battery capacity can still support all three carriers if compensated by higher electrolyzer and hydrogen storage capacities. However, when battery capacity is high, methanol pathways are favored: ammonia and methane together account for only about a quarter of the feasible import configurations (26\%). Adding one more decision layer, hydrogen storage, reinforces this pattern. High hydrogen storage capacity, combined with high battery capacity and low wind power, eliminates ammonia and methane entirely, leaving methanol as the sole remaining option. Such systems rely on PV-dominated electricity, extensive storage, and smaller electrolyzers operating steadily.

Taken together, \autoref{fig:fuel_classification_dt} reveals three main design archetypes. Wind-dominated designs tend to consider ammonia or methane as an energy carrier and operate with low PV and storage capacities. Mixed wind–PV systems accommodate all three carriers and offer the greatest design flexibility, with methanol pathways limited to configurations with high storage. Finally, PV-dominated systems with large storage capacities are mainly methanol-based. The tree also highlights the strong coupling between ammonia and methane, which share similar branches and design conditions. Methanol, in contrast, stands out with a distinct design archetype characterized by low wind capacity and high PV and storage requirements.

\subsubsection{Carrier-specific decisions}
\label{sec:carrier_specific_analysis}

Following the interpretation of the near-optimal space for all non-hydrogen carriers combined, in this section, we will dive deeper into the decisions related to a specific carrier pathway. We start by discussing the hydrogen pathway, followed by the methanol import pathway. The decision trees for ammonia and methane, which are similar to that of hydrogen, are provided in Supplementary Information.

\begin{figure}[h!]
    \centering
    \includegraphics[width=1\linewidth]{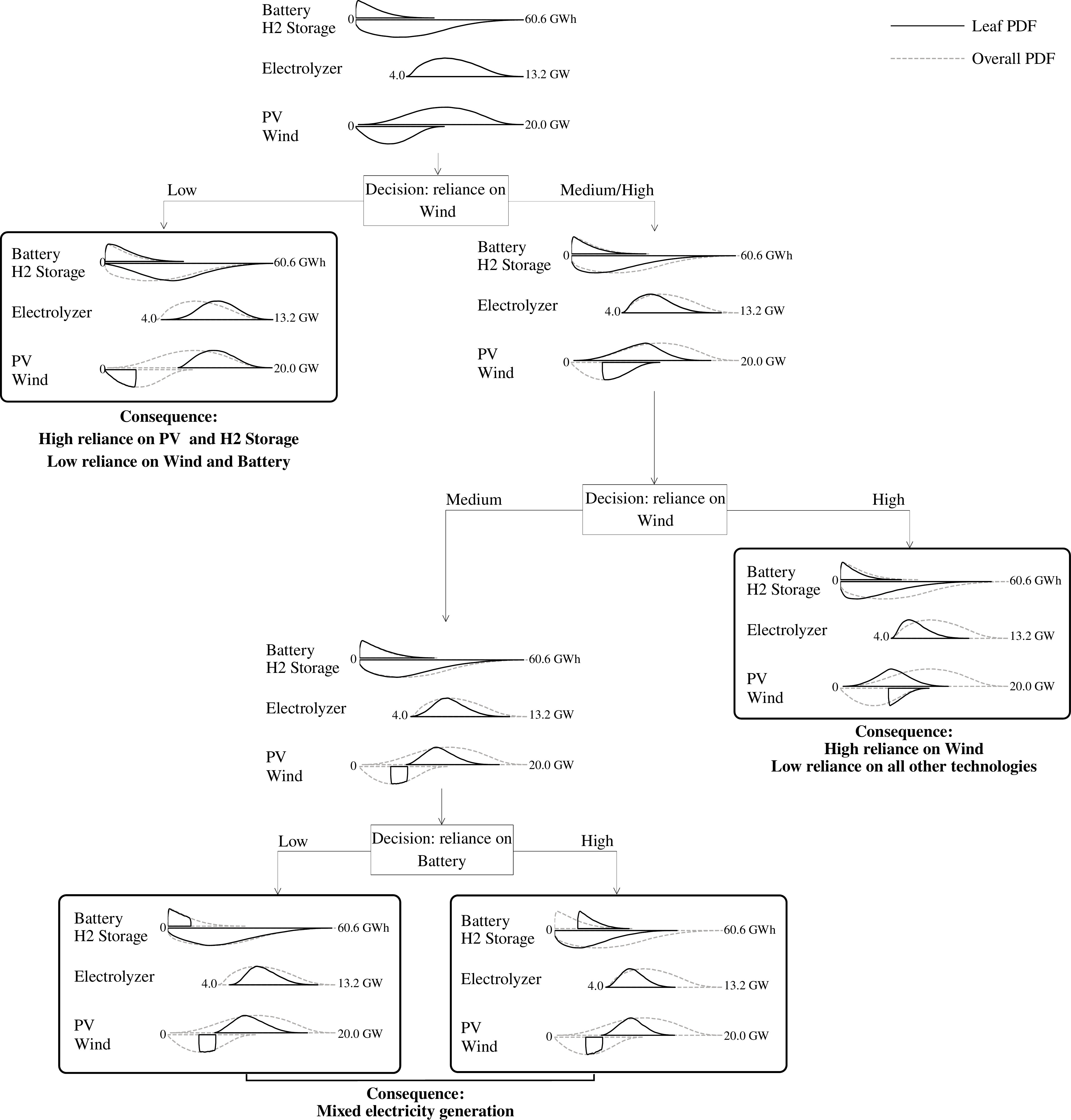}
\caption{Decision tree for the hydrogen pathway showing how key design decisions affect the rest of the system. The first decision concerns wind capacity: low wind capacity leads to higher PV generation, larger electrolyzer capacity, and greater hydrogen storage, whereas high wind capacity reduces the need for these technologies. In between, a balanced mix of PV and wind allows smaller electrolyzer and storage capacities when battery capacity is high.}
    \label{fig:single_fuel_dt}
\end{figure}

For the hydrogen case, the decision tree shows that the most influential decisions concern wind power capacity, followed by battery capacity (\autoref{fig:single_fuel_dt}). These two variables drive most of the variation in system design and determine how other technologies are sized to remain within the near-optimal cost range.

The first key decision concerns wind capacity. Choosing low wind capacity requires decision-makers to deploy higher PV generation, larger electrolyzer capacity, and greater hydrogen storage (leftmost branch). In these cases, PV capacity lies toward the upper end of its range (8.8~GW to 20~GW). The electrolyzer and hydrogen storage capacities follow similar upward trends, with most near-optimal designs clustering at higher values. The flexibility regarding battery capacity, however, remains largely unchanged.

When wind capacity is high (rightmost branch), the pattern reverses. These configurations require less PV installation, smaller electrolyzers, and reduced hydrogen storage. Both PV and electrolyzer capacities fall within the lower half of their respective ranges, while hydrogen storage also decreases, as indicated by the skewness of its density distribution.

Between these two extremes, the model identifies a middle case where wind and PV capacities are more balanced. In these configurations, battery capacity plays a secondary role: low battery capacity has little impact, whereas higher battery capacity reduces the need for electrolyzer and hydrogen storage.

Although configurations that exclude either wind or PV power entirely are technically feasible, they are rare within the near-optimal set and limit flexibility for other technologies. In contrast, systems with low or no storage capacities are more common, demonstrating that storage can be constrained without significantly deviating from the near-optimal range. Overall, no technology is strictly required, but each decision influences the others. In practice, reducing one technology, such as wind, PV, or storage, necessitates compensating increases elsewhere, underscoring the strong interdependence among all components.

\begin{figure}[h!]
    \centering
    \includegraphics[width=0.8\linewidth]{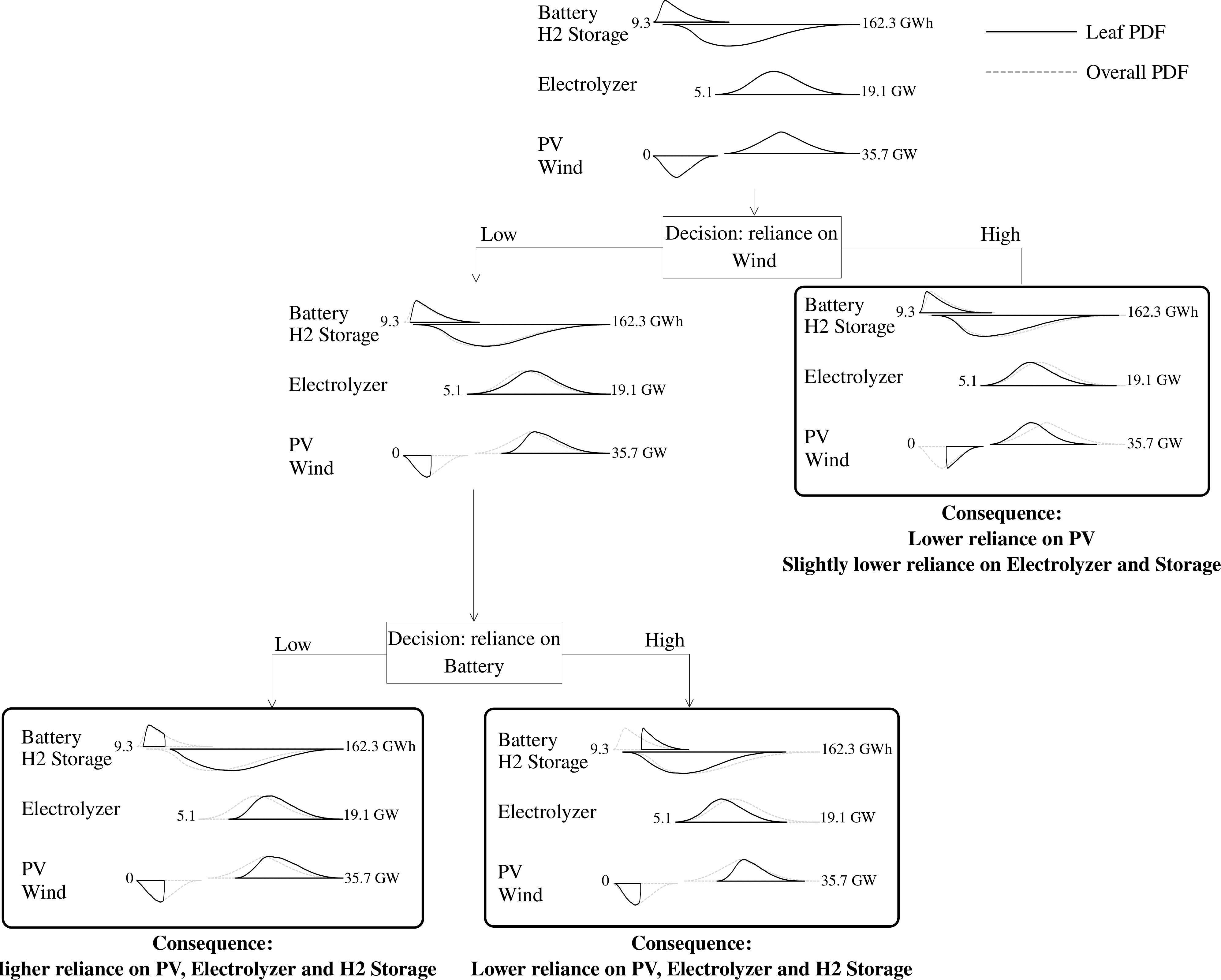}
    \caption{Decision tree for the methanol pathway showing how key design choices shape the near-optimal system configurations. 
    The first node separates designs by wind power capacity, which, unlike for other carriers, has little effect on PV, electrolyzer, or storage capacities. The second node splits by battery capacity, where higher battery capacity slightly reduces PV, electrolyzer, and hydrogen storage requirements.}
    \label{fig:meoh_dt}
\end{figure}

For the methanol case (\autoref{fig:meoh_dt}), the decision patterns differ from those of the other carriers. As discussed in \autoref{subsec:near_optimum_space}, PV capacity is never excluded, and storage capacities, both battery and hydrogen, are generally higher. Although the feasible range of storage capacities extends to low values, the associated density is very small, indicating that most near-optimal methanol systems rely on high storage levels.

The first node splits the solutions according to wind power capacity. Unlike the other carriers, this decision has only a minor influence on the rest of the system. Moving from the left branch (low wind) to the right branch (high wind) does not significantly reduce the need for PV, storage, or electrolyzer capacities. The density plots remain nearly identical across branches, with only a slight decrease in PV and electrolyzer capacities at higher wind levels.

The second node concerns battery capacity. Along the upper branch, where battery capacity is high, the system moves towards a slightly lower electrolyzer, hydrogen storage, and PV capacities archetype. Along the lower branch, where battery capacity is low, these capacities slightly increase. However, the changes remain small compared to those observed for the other carriers, indicating limited interdependence among design variables.

Overall, the methanol pathway shows weaker trade-offs in comparison to the others. Nearly all near-optimal designs feature relatively high PV generation and hydrogen storage capacities, regardless of wind or battery choices. This suggests that methanol systems depend on stable, solar-driven electricity and substantial storage to sustain continuous operation, while variations in wind and battery capacity play only a secondary role.

\section{Discussion}
\label{sec:discussion}
Stakeholders rarely pursue purely techno-economic optimality: instead, they balance costs with risk, institutional feasibility, and other practical considerations~\cite{Brill1979}. Hence, it is essential to move beyond single cost-optimal solutions and explore near-optimal alternatives that remain economically comparable but differ in their technical and institutional characteristics. Equally important is to organize these alternatives in an interpretable way to ensure transparent communication of insights. In this regard, applying interpretable machine-learning methods to MGA results reveals decision patterns within the near-optimal space, clarifying the trade-offs involved in energy investment decisions and showing how designs can adapt when circumstances change.

In our analysis of non-hydrogen carriers, the wind-dominated pathways, which favor hydrogen production and transport chains based on ammonia or methane, align with a preference for proven technologies and, to some extent, more predictable operations. These designs are simpler to coordinate and can fit within existing industrial and regulatory frameworks~\cite{cao2025comprehensive}. Ammonia benefits from a mature global shipping network and well-defined safety standards~\cite{Cho2024IJHE}, while synthetic methane can build on existing gas infrastructure and handling experience~\cite{Carels2024ECM}. These pathways are easier to finance and approve because they build on familiar assets, limiting the use of novel technologies and lowering the associated risks. For wind-rich coastal regions such as Morocco, these systems are particularly attractive from a technical and legislative standpoint, making implementation easier.

In contrast, PV-dominated configurations are best with methanol pathways. The liquid nature of methanol facilitates storage and transportation and may improve public acceptance compared to hydrogen or methane~\cite{reddy2023sustainable}. However, these systems depend on reliable CO$_2$ sourcing, certification mechanisms, and regulatory support for carbon management. They require coordination between CO$_2$ capture, synthesis, reforming, and verification frameworks, which are still maturing~\cite{McLaughlin2023RSER}. As a result, methanol-based pathways are technologically feasible but still lack full maturity in terms of institutional readiness and policy stability.

Between these two poles, mixed wind–PV configurations occupy a valuable middle ground in the near-optimal landscape. They remain economically competitive while keeping all carrier options open, offering flexibility to adapt to evolving policies, as hydrogen and carrier supply chains are likely to evolve under uncertain financing and regulatory conditions~\cite{Odenweller2022NatureEnergy}. 

This discussion applies only within a non-hydrogen carrier framework. In other words, it holds only when hydrogen is not considered as a transport carrier. If hydrogen is included, the discussion changes completely, since its cost advantage over the other carriers must also be taken into account.

Going further, our analysis shows that even when a specific carrier is selected, which already constrains the design space, there remains great flexibility. For instance, while the general trend in \autoref{sec:carrier_classification} suggests that ammonia and methane based systems are favored under high wind capacity, a closer look at each carrier’s decision tree in \autoref{sec:carrier_specific_analysis} reveals that low-wind configurations can still be feasible and cost-competitive. This flexibility becomes especially relevant once projects move into construction. If market conditions or supply-chain shocks render one direction impractical, decision-makers have other alternative configurations already identified within the same cost range. For example, if disruptions in the supply of rare earth metals delay wind turbine deployment~\cite{DEPRAITER2025108496}, \autoref{sec:carrier_specific_analysis} shows that systems can compensate by relying more heavily on PV and storage. In contrast, if safety concerns or environmental impacts limit the large scale deployment of lithium-based batteries~\cite{CHEN202183, LAVISSE2024766}, designs may shift toward wind-oriented configurations, reducing the dependence on storage.

The main takeaway is that near-optimal exploration reveals several cost-comparable yet structurally distinct ways to produce hydrogen. No single technology mix is essential. Wind, PV, and mixed systems each occupy their own portion of the near-optimal space, offering viable options that depend on context, infrastructure, and policy priorities. This diversity shows that hydrogen supply chains can take many equally viable forms within similar cost bounds. Importantly, uncovering this range is only possible through near-optimal exploration using MGA, which is therefore essential for identifying the full spectrum of practical and policy-relevant alternatives for hydrogen imports from renewable-rich locations to high-demand areas.

Despite the insights this work provides, it still has several limitations that should guide future research. First, the model used here is linear, which neglects the dynamic behavior of individual components (e.g., electrolyzer load dynamics) and, importantly in the context of MGA, economies of scale. Given the high flexibility in installed capacities, ranging from none to maximum capacity, assuming a constant unit cost across the entire range is not the most accurate approach. A more realistic representation would involve higher unit costs at the lower end of the range. Second, all parameters in this study are deterministic. While this approach highlights the design implications of decisions, it does not capture the influence of input variability, such as changes in technology costs, efficiencies, or renewable availability, on outcomes~\cite{mavromatidis2018review, NEUMANN2023106702}. Future work will extend the framework with parametric uncertainty analysis to assess the robustness of near-optimal configurations under uncertain techno-economic and operational conditions. Third, the definition of near-optimality depends on the allowed cost slack in MGA. The 10\% slack used here is common~\cite{decarolis2011}, yet subjective: reducing it would narrow the design space and suppress diversity. Moreover, the MGA approach used in this work limits the number of decision variables, leaving out other design parameters that may also be of interest. 

\section{Conclusion}
\label{sec:conclusion}

In this work, we explored the near-optimal space of hydrogen import pathways, considering diversity in technology capacities and carriers, and streamlined it into key actionable decisions. This approach overcomes the limitation of relying solely on the single mathematically optimized solutions, which are vulnerable to non-quantifiable constraints that can restrict the decision space, making the optimal solution infeasible or undesirable.

The near-optimal space revealed a vast design range, in which technologies (except the electrolyzer) can be entirely excluded while maintaining a cost increase below 10\% of the optimal LCOH. Given the large number of near-optimal solutions, extracting clear information for decision-makers is challenging. Therefore, a combination of clustering and decision trees was employed to streamline these outputs into actionable insights. The results revealed that ammonia and methane share similar design characteristics, relying more on wind and less on other technologies, whereas methanol systems rely more on PV and storage and less on wind. For a specific carrier, three main design archetypes were identified: a PV-dominated archetype with high electrolyzer and hydrogen storage usage and low wind usage, a wind-dominated archetype with low use of all other technologies, resulting in storage-free designs, and a mixed electricity generation archetype, where battery capacity influences electrolyzer design, with higher battery capacity leading to lower electrolyzer capacity.

The findings of this work can support decision-makers in understanding the design options as well as their consequences. These insights not only present the various alternatives but also help extract clear decision-making rules and identify trade-offs, such as the need for high wind capacities when high storage capacities are not desired or feasible. This can assist in aligning design choices with site characteristics and facilitate clearer communication of project attributes to stakeholders from the early stages of planning.

Future work will focus on applying this method to broader applications, such as transition models with pathway formulations, as well as improving the MGA method to include parametric uncertainty and to overcome dimensional limitations.

\section*{Supplementary information}
Supplementary information is available for this paper.

\section*{CRediT authorship contribution statement}
\noindent
\textbf{Mahdi Kchaou:} Conceptualization, Data Curation, Formal Analysis, Investigation, Methodology, Software, Validation, Visualization, Writing – Original Draft, Writing – Review \& Editing. \textbf{Francesco Contino:} Conceptualization, Visualization, Writing – Review \& Editing. \textbf{Diederik Coppitters:} Conceptualization, Formal Analysis, Methodology, Supervision, Validation, Visualization, Writing – Original Draft, Writing – Review \& Editing.

\section*{Declaration of competing interest}
\noindent All authors have no competing interests.

\section*{Acknowledgments}
\noindent
This work was supported by the Wallonie-Bruxelles International - WBI (M.K); and the Fonds de la Recherche Scientifique - FNRS [CR~40016260] (D.C).


\bibliography{mybibfile}

\end{document}